\documentclass[preprint]{vgtc}               

\graphicspath{{figures/}{pictures/}{images/}{./}} 

\usepackage{times}                     

\usepackage{tabu}                      
\usepackage{booktabs}                  
\usepackage{lipsum}                    
\usepackage{mwe}                       

\usepackage{amssymb}
\usepackage{mathptmx}                  

\onlineid{0}

\vgtccategory{Research}

\vgtcinsertpkg

\title{Visual Compositional Data Analytics for Spatial Transcriptomics}

\author{David H\"{a}gele\thanks{e-mail: david.haegele@visus.uni-stuttgart.de}\\ %
        \scriptsize University of Stuttgart %
\and Yuxuan Tang\thanks{e-mail: st189806@stud.uni-stuttgart.de}\\ %
     \scriptsize University of Stuttgart %
\and Daniel Weiskopf\thanks{e-mail: daniel.weiskopf@visus.uni-stuttgart.de}\\ %
     \scriptsize University of Stuttgart}

\teaser{
  \centering
  \includegraphics[width=\linewidth]{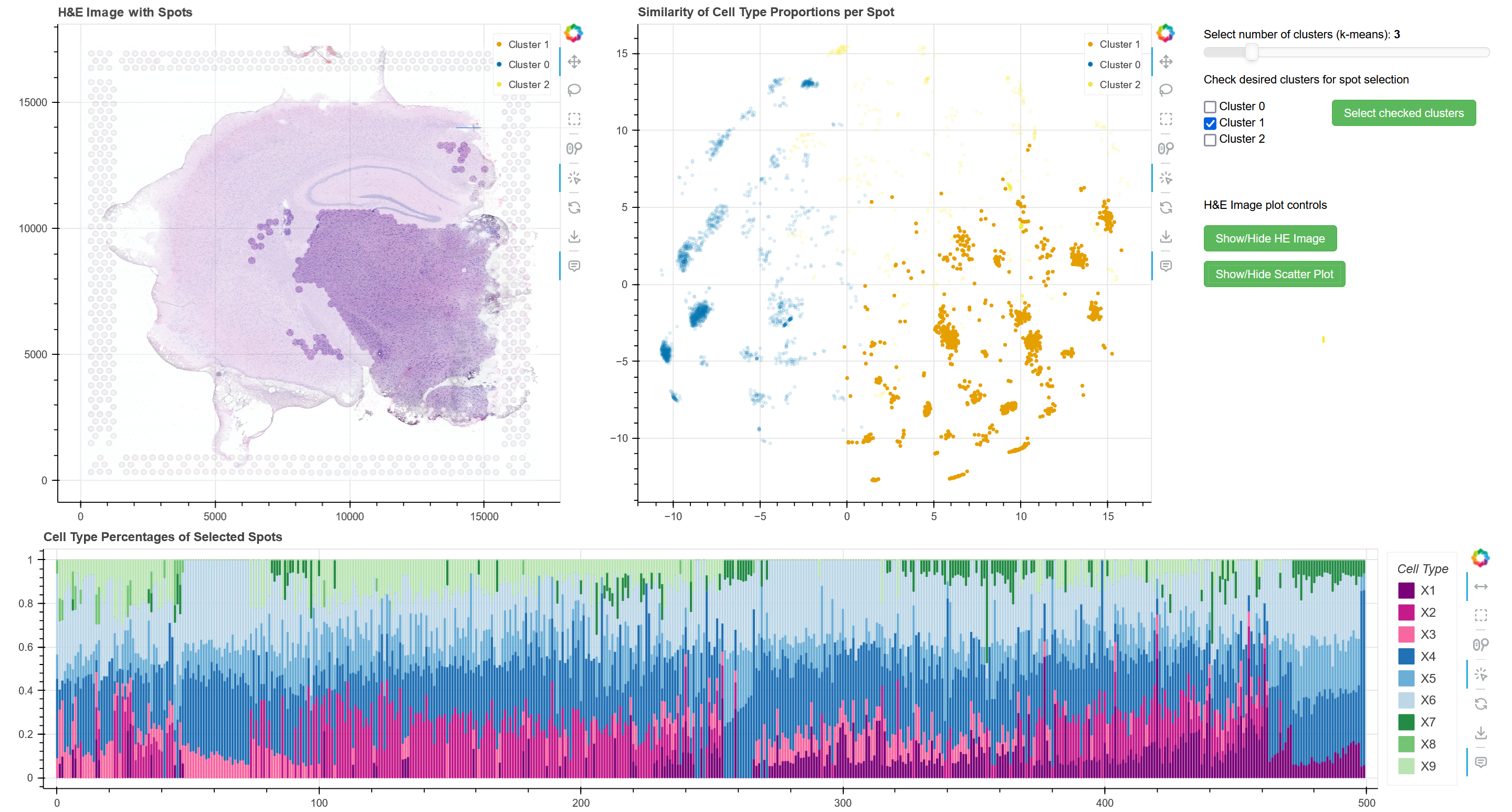}
  \caption{Our visual analytics tool for exploring the cell type proportions detected on the tissue.
  At the upper left, the histological image with highlightable spots. Next to it, a dimensionality reduction of the individual cell type proportions detected at the spots. On the right, controls for selecting clusters from $k$-means clustering and hiding/showing the spots/image in the left view.
  At the bottom, a stacked bar chart showing the cell type proportions of selected spots.}
  \label{fig:teaser}
}

\abstract{
    For the \emph{Bio+Med-Vis Challenge 2024}, we propose a visual analytics system as a redesign for the scatter pie chart visualization of cell type proportions of spatial transcriptomics data.
    Our design uses three linked views: a view of the histological image of the tissue, a stacked bar chart showing cell type proportions of the spots, and a scatter plot showing a dimensionality reduction of the multivariate proportions.
    Furthermore, we apply a compositional data analysis framework, the Aitchison geometry, to the proportions for dimensionality reduction and $k$-means clustering.  
    Leveraging brushing and linking, the system allows one to explore and uncover patterns in the cell type mixtures and relate them to their spatial locations on the cellular tissue.
    This redesign shifts the pattern recognition workload from the human visual system to computational methods commonly used in visual analytics.
    We provide the code and setup instructions of our visual analytics system on GitHub.\footnote{\url{https://github.com/UniStuttgart-VISUS/va-for-spatial-transcriptomics}}
} 


\begin{document}


\firstsection{Introduction}

\maketitle

Spatial transcriptomics technology enables capturing gene expression in the spatial context of cellular tissue~\cite{spatialtranscriptomics}.
Per \emph{spot} (capture location on the tissue), a mixture of different cell types is detected. 
Ultimately, single cell types would be desirable but are hard to capture due to limited resolution~\cite{STsummary}.
However, recovering cell type proportions per spot is possible~\cite{celltypedeconvolve}, leading to compositional data that can be analyzed in relation to the spatial context of the tissue.

Previous work has used glyph-based representation of the cell type proportions, such as small pie charts, superimposed on a corresponding histological image of the tissue.
Such visualizations leverage human perception for identifying patterns, such as areas of similar data items.
However, the use of glyphs, especially pie chart glyphs, has its limitations and problems.
The readability of pie charts depends on the choice of colors, which becomes difficult for many pie slices and for color blindness support.
Smaller variations in the data are difficult to spot since humans perform worse at comparing areas compared to other retinal variables, such as length or position.
Superimposed glyphs in general obstruct the view onto the image, thus, hindering the examination of the corresponding tissue spots.

Our solution for this visualization redesign challenge shifts away from glyphs and employs a visual analytics approach with several linked views.
Instead of primarily relying on human perception, we leverage computational methods to recognize patterns in the different cell type proportions.
Specifically, we utilize the Aitchison geometry intended for compositional data~\cite{compositionaldata} in conjunction with clustering and dimensionality reduction.

\begin{figure*}[t!]
    \centering
    \includegraphics[width=\linewidth]{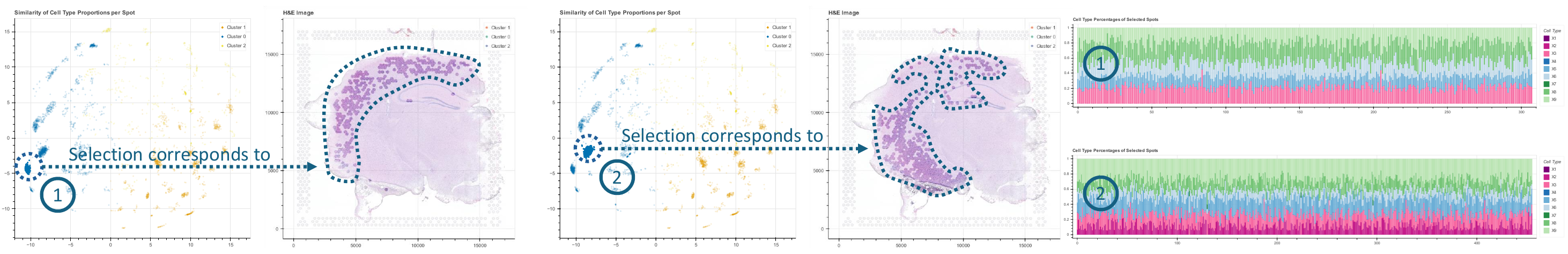}
    \caption{Selecting different blobs in the dimensionality reduction view highlights corresponding areas that exhibit different proportion patterns.}
    \label{fig:selections}
\end{figure*}

\section{Design}\label{sec:design}
Our visual analytics system is intended for spatial transcriptomics data, comprising the following:
\begin{itemize}\itemsep-.5em
    \item A set of cell type proportions $p_i \in \Delta^{d-1}$, where $\Delta^{d-1}$ is the $(d-1)$-standard simplex with $d$ non-negative components that sum up to $1$.
    \item Corresponding spatial locations $x_i \in \mathbb{R}^2$ (spots) on the tissue to which the proportions are assigned.
    \item A histological image of the corresponding tissue.
\end{itemize}
\Cref{fig:teaser} shows our system, which consists of three separate linked views.
We show the spatial information, i.e., the histological image and the spots $x_i$ in one view. The spots can be displayed as superimposed dots to mark their locations, but also a selection of spots can be used to create an image mask so that the tissue can be observed without occlusion (dark purple parts in~\Cref{fig:teaser}).

The cell type proportions $p_i$ are visualized by a stacked bar chart.
Since the number of spots $n$ is larger than a regular monitor's horizontal resolution ($n>2000$), not all proportions can be displayed at once.
Instead, we limit the number of bars to $500$ and only show a subset.
When a selection of spots is triggered (e.g., from another view), the bar chart is updated to only show the corresponding proportions.

In a scatter plot, we show a dimensionality reduction of the \mbox{$d$-dimensional} proportions.
For any automated analysis or transformation of the data $p_i$, it is crucial that it is treated as compositional to account for the differences in ratios rather than values~\cite{gloor2017compositional}.
Therefore, we apply a center log-ratio transformation to the $p_i$ so that regular linear algebra operations are done in Aitchison geometry, which guarantees that we stay within the simplex $\Delta^{d-1}$.
Then, a 2D projection is obtained using principal component analysis.
We also use the projection along the first principal component as ordering for the bars of the bar chart, which provides better visual consistency for many bars since the bars become roughly ordered by similarity.
A subset in case of too many data items can therefore be created similar to nearest neighbor image downscaling.

The same center log-ratio coordinates are used to apply $k$-means clustering.
The cluster assignment of the corresponding spots is shown by color-coding in the dimensionality reduction view and the image view (when spot markers are enabled).
A control panel in our system allows selecting $k$ and triggering spot selection of specific clusters.

Our system strongly relies on interactive exploration through brushing of data items in any of the views.
For example, selecting spots in the image view provides an overview of the cell type proportions of the selected area.
Picking items in the dimensionality reduction view or by cluster allows us to identify similar proportions, their corresponding locations on the tissue, and identify the dominant cell types.
\Cref{fig:selections} shows an example of such a selection and how the corresponding patterns differ.
From the stacked bar chart, specific data items of interest can be selected based on their proportion and pattern. 
Furthermore, a linked table view lists all the data items and allows for sorting and filtering according to specific cell type percentage.

\section{Limitations and Outlook}
Our solution only uses the data described in \Cref{sec:design}, but the provided challenge data set contains more data about the specific genes and their relation to the cell types, which cannot be analyzed currently.
We excluded this data because the scatter pie plot to redesign did not encode such data either.
However, our system is quite extendable. Therefore, supporting additional data would be easy.

While stacked bar charts are generally prone to be misinterpreted by laymen, our system is designed for scientists who we expect to be familiar with this chart type. 
However, the view does not scale well with the total number of data items and therefore cannot show the entirety of the data.
This limitation is mitigated by the dimensionality reduction view, which provides an abstraction of the data items based on their similarity.
Our solution to show a subset when there are too many data items was simple to implement but misses to represent discarded items and lacks the possibility for selection of them.
This should be replaced with a more sophisticated summarization approach in the future.


The presented system is in a  prototypical state, and many possible interaction and encoding techniques that would facilitate exploration are not implemented, e.g., a hovering method that would allow locating corresponding items in the different views quickly, or a feature that would highlight other spots with similar proportions were planned but not realized. The most limiting factor of this design is the lack of requirements and evaluation from domain scientists regarding their data analysis.
Nevertheless, our system provides good explorability of the data
and possibilities to relate cell type mixtures to the spatial context of the tissue.

\acknowledgments{
This work was funded by the Deutsche Forschungsgemeinschaft (DFG, German Research Foundation)---Project ID 251654672---TRR 161 (Project A01).
}

\bibliographystyle{abbrv-doi-hyperref}

\bibliography{template}
\end{document}